\documentclass[apj,numberedappendix]{emulateapj}

\def\gtorder{\mathrel{\raise.3ex\hbox{$>$}\mkern-14mu
             \lower0.6ex\hbox{$\sim$}}}
\def\ltorder{\mathrel{\raise.3ex\hbox{$<$}\mkern-14mu
             \lower0.6ex\hbox{$\sim$}}}

\slugcomment{Draft of \today}
\shorttitle{The Arecibo Fast Radio Burst}
\shortauthors{Kulkarni, Ofek \&\ Neill}

\begin{document}

\title{The Arecibo Fast Radio Burst: Dense Circum-burst Medium}
\author{
S. R. Kulkarni\altaffilmark{1},
E. O. Ofek\altaffilmark{2} \& J. D. Neill\altaffilmark{3} 
}
\altaffiltext{1}{Caltech Optical Observatories 249-17 Caltech, Pasadena, CA 91125, USA}
\altaffiltext{2}{Department of Particle Physics \&\ Astrophysics, Weizmann Institute of
Science, Rehovot 76100, Israel}
\altaffiltext{3}{Space Radiation Laboratory 290-17, Caltech,  Pasadena, CA 91125, USA}

\begin{abstract}

The nature of fast radio bursts (FRB) showing a single dispersed
few-ms-width pulse, with dispersion measure (DM) in excess of the Galactic
value has been extensively debated.  Here we investigate FRB\,121102, the first FRB detected
at the 305-m Arecibo radio telescope and remarkable for its unusually
large spectral index. After extensive explorations of options we conclude that
the spectral index is caused by a nebula with free-free absorption. 
Observations (or lack thereof) show that the nebula is located beyond
the Milky Way. 
We conclude that FRBs are of extra-galactic origin and that they arise
in dense star-forming regions. The challenge with extra-galactic
models is the the high volumetric rate of FRBs. This high
rate allows us to eliminate all
forms of catastrophic stellar death as a progenitor.  Hyper-giant flares
from young magnetars emerge as the most likely progenitors.  A number of
consequences follow from this hypothesis: (i) FRB models which posit a
purely intergalactic origin can be safely ignored.  (ii) The rich ISM
environment of young magnetars can result in significant contribution to
DM, Rotation Measure (RM) and in some cases to significant free-free
optical depth.  (iii)  The star-forming regions in the host galaxies can
contribute significantly to the DM. Including this contribution reduces
the inferred distances to FRBs and correspondingly increases the
volumetric rate of FRBs (and, in turn, may require that giant flares
can also produce FRBs).
(iv) FRBs are likely to be suppressed at lower frequencies.
Conversely, searching for FRBs at higher frequencies (2--5\,GHz) would be
attractive.  (v) The blast wave which produces the radio emission can
undergo rapid deceleration if the circum-burst medium is dense (as maybe
the case for FRB\,121102), leading to X-ray, radio and possibly
$\gamma$-ray emission.  (vi) Galaxies with high star formation rate host
will have a higher FRB rate. 
However, such FRBs will have differing
DMs owing to differing local contributions. (vi)
The DM and RM  of FRBs will prove to be noisy probes of the intergalactic
medium (density, magnetic field) and cosmography.

\end{abstract}


\section{Background}
\label{sec:Background}

Fast Radio Bursts (FRBs) were first identified in an archival
analysis of pulsar data obtained with a multi-beam receiver at the
64-m radio telescope of the Parkes Observatory and operating in the
1.4\,GHz band \citep{lbm+07}.  The few millisecond wide burst showed
a quadratic frequency dependent arrival time similar to pulsar
signals traversing the Galactic ionized interstellar medium (ISM).
The inferred dispersion measure (DM) was greatly in excess of that
expected from a Galactic source.  A simple interpretation was that
the burst of radiation came from another galaxy (or was located in
intergalactic space) with the excess DM due to electrons in the
intergalactic medium (IGM).

However, subsequently, ``perytons" -- radio bursts, but appearing
in many (all) beams -- were discovered. Telescopes are focused for
sources at infinity and so celestial sources typically occur in
only one beam.\footnote{FRB\,010724, the super-bright event discovered
by \citet{lbm+07}, occurred in several beams.  At times some FRBs,
if located in between the sky positions of two  beams, are found
in adjacent beams.} In contrast, local sources appear in many beams.
Thus, perytons cast doubts about the celestial origin of FRBs.
However, recently perytons were shown to emanate from a microwave
oven at the Parkes Observatory \citep{peb+15}. With perytons firmly
established to be of domestic origin the standing of FRBs has risen.

For FRB\,010724, Kulkarni {\it et al.} (2014; hereafter,
K14)\nocite{kon+14} considered the possibility of an intervening
Galactic nebula to account for the DM in excess of that provided
by the Galactic ISM.  Such a nebula would shine via emission
(recombination lines) and also absorb radio signals (via free-free
absorption).  No plausible nebula was identified within the
localization region of FRB\,010724.

To date there are fifteen FRBs reported  from the Parkes Observatory
\citep{kp14,rsj15,cpk+15}). This number would be doubled by including
FRBs discussed in the hallways of astronomy departments.  The typical
peak flux density, $S_p$, at the fiducial frequency, $\nu_0=1.4\,$GHz,
is between 0.5\,Jy to 2\,Jy. The durations\footnote{Note that there
is loss of sensitivity for  sub-ms duration bursts} ($\Delta\tau$)
range from  1\,ms to 10\,ms.  FRB\,010724 is the brightest to date
with a fluence of $\mathcal{F}=S_p\Delta t$ of several hundred
Jy\,ms and also has one of the lowest dispersion measures,
375\,cm$^{-3}$\,pc. The highest reported DM  is about 1629\,cm$^{-3}$\,pc
\citep{cpk+15}.  The overall daily all-sky rate is 2500 events with
$\mathcal{F}\ge 2\,$Jy\,ms \citep{kp14} and $6^{+4}_{-3}\times 10^3$
for the entire detected Parkes sample \citep{cpk+15}.

The extra-galactic origin for FRBs, whilst very appealing,\footnote{The
diagnostic value of readily  detectable millisecond pulses originating
from cosmological distances is immense, perhaps even transformational.}
has two great challenges (see K14 for further details). First FRBs,
if extragalactic, would be as brilliant as the the nanosecond pulses
from pulsars except that FRBs last for a few milliseconds. The
energy radiated in the radio alone would be an impressive $10^{40}\,$erg.
Next, assuming most of the inferred DM is due to electrons in the
IGM, the volumetric rate of FRBs is an impressive ten percent  of
the core-collapse supernova rate. A key requirement for any model
of FRBs is that the radio pulse has to be able to propagate.  This
means that the circum-burst density must be low enough that the
resulting plasma frequency is lower than $\nu_0$. Next, the optical
depth to free-free opacity should also not be prohibitively large.
These requirements rule out all supernovae models (see K14).  So
the progenitors of FRBs  come from (i) a  new channel of stellar
death or  (ii) an entirely new channel of non-stellar phenomenon
or  (ii) some sort of repeating phenomenon.  The first two possibilities
would be revolutionary.  The third possibility, whilst less alarming,
would have to rise to the challenges posed by the duration and
energetics of FRBs.

\section{The Arecibo Fast Radio Burst}
\label{sec:AFRB}

FRB\,121102 was found in an analysis  of archival data (\citealt{sch+14};
hereafter, S14) obtained from the seven-beam receiver located at
the Gregorian focus of the Arecibo 305-m radio telescope and operating
in the 1225--1525\,MHz band \citep{cfl+06}. This event is noteworthy
for being the first FRB detected outside the Parkes Observatory.

The specific pulsar survey targeted the Galactic anti-center and
lasted 11.8\,d. The burst was detected in beam number 4 of the
multi-beam system with an SNR of 11 (integration time of 3\,ms).
The sky position of the center of beam 4 is $l=174^\circ.95$ and
$b=-0^\circ.223$ which corresponds to RA=05h32m9.6s \&\
Declination=33d05m13s (J2000).  No bursts were seen during subsequent
observations of the same region of the sky.


The event lasted  about 3\,ms and exhibited a frequency-dependent
arrival time, $t_a\propto \nu^m$ with $m=-2.01\pm 0.05$, consistent
with that expected for a signal propagating through cold plasma.
The apparent DM is $557.4\pm 2$\,cm$^{-3}$\,pc.  The pulse width
appears to be independent of the frequency.  After accounting for
the spectrometer resolution, S14 determine the intrinsic pulse width
is $\Delta\tau_d<1\,$ms at  $\nu_0=1.4$\,GHz.  The best fit spectral
index\footnote{Here, $S(\nu)$, the flux density at frequency $\nu$
is modeled as $S(\nu)\propto \nu^\beta$.} is positive, $7<\beta<11$.
This is a remarkable spectral index especially for coherent radio
emission.\footnote{On good grounds we expect the fundamental mechanism
of FRBs is some sort of coherent emission.}

The inference of the peak flux density, $S_p$, depends on the sky
location of FRB\,121102 with respect to the principal axis of beam
4.  The gain of a feed is maximum along the principal axis and
decreases with increasing angle. Sources can be detected outside
the nominal ``beam at half maximum" provided they are bright.  The
peak antenna temperature of FRB\,121102 is about 0.28\,K. If the
event is located in the primary beam of beam 4 then $S_p=0.04\,$Jy
which would make it the faintest FRB to date.  On the other hand,
S14, noting the unusual spectral index, argue the source was detected
in the side lobe. If so, $S_p \approx 0.4\,$Jy, not too different
from those of FRBs detected at Parkes.  In short,  
the fluence of FRB\,121102 at the fiducial frequency,
$\nu_0$, can be as small as  0.12\,Jy\,ms or as large as 1.2\,Jy\,ms.

FRB\,121102 is located close to the Galactic Equator ($b\approx
-0.22^\circ$) and towards the anti-center region ($l\approx
175^\circ$).  According to S14, the Galactic ISM, if integrated to
the edge of the Galaxy, would contribute a dispersion measure of
about 188\,cm$^{-3}$\,pc. If FRB\,12110 is extragalactic then the
excess DM is about 370\,cm$^{-3}$\,pc.

S14 considered terrestrial origins as well as a Galactic origin
(Rotating Radio Transients). They disfavor both these two possibilities.
Noting that FRB\,121102 shared many similarities with events  found
at the Parkes Observatory \citep{lbm+07,tsb+13} S14 enthusiastically
joined their colleagues at Parkes in advocating an extra-galactic
origin for FRB\,121102.

\subsection{Why is FRB\,121102 interesting?}

In the short history of FRBs, FRB\,121102 occupies a special position.
We have already noted the remarkable spectral index. Next, given
the detection of this event in a single Arecibo beam, is that
FRB\,121102 has to lie beyond the Fresnel length or about 100\,km
away from the telescope. These two observations provide unique
insights into the origin of FRBs and provide the motivation for
this paper.

The structure of the paper is as follows.  The two unique attributes
of FRB\,121102 are discussed in \S\ref{sec:MinimumDistance} and
\S\ref{sec:AmazingSpectralIndex}.  We explore several possibilities
to account for the spectral index and conclude that the large
positive spectral index is due free-free absorption arising in an
intervening ionized clump (nebula). The locale of this putative
nebula (galactic, extra-galactic, circum-FRB) are explored in
\S\ref{sec:InterveningNebula}--\S\ref{sec:AnExtraGalacticOrigin}.
In  \S\ref{sec:GFFRB} we present our ``best-buy" model,
namely that FRBs are powered by (hyper) giant flares from young
Soft Gamma-ray Repeaters (SGRs). This model accounts for the apparent
fecundity of FRBs as well as provides a most natural explanation
for the peculiar spectral index. We end the paper by discussing the
considerable ramifications of this model (\S\ref{sec:Ramifications}).

\section{The Minimum Distance to FRB\,121102}
\label{sec:MinimumDistance}

\begin{figure}[htbp] 
   \centering
   \includegraphics[width=7.5cm]{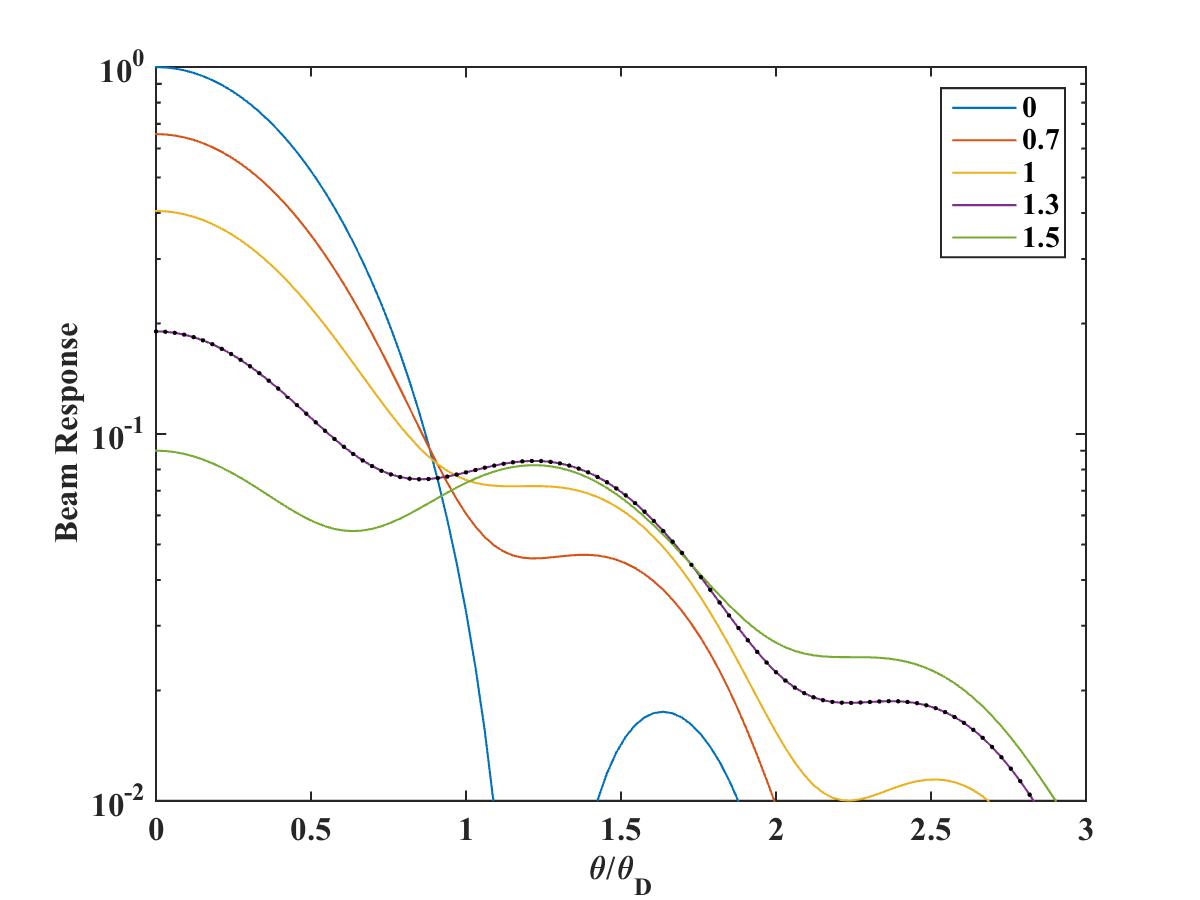} 
   \caption{\small The beam response of a circular aperture illuminated
   uniformly for a point source located at a distance $d=a_F/n_F$
   as a function of the Fresnel number, $n_F$ (whose values can be
   found in the legend box).  The angular offset is units of the
   diffraction angle, $\theta_D=\lambda/\mathcal{D}$ where $\mathcal{D}$
   is the diameter of the aperture. The response is normalized to
   unity for point source which is along the principal axis of the
   aperture and located at infinity.
	}
\label{fig:FresnelResponse} 
\end{figure}

Radio telescopes are tuned to receive signals from sources very far
away (Fraunhofer regime). Thus, ``nearby" (local) sources are
expected to be out of focus and appear in several beams (e.g., perytons). The ``single
beam" criterion is usually interpreted to mean that the event is
at a great distance from the telescope. The Fresnel length scale,
$a_F$, separates nearby sources from very distant sources. It is
given by\footnote{Note that the expression for $a_F$ in K14 is off
by a factor of 4.  The Fresnel scale  is the ratio of the square
of the radius to the wavelength.}
	\begin{equation}
		a_F=\frac{\mathcal{D}^2}{4\lambda}
	\end{equation}
where $\mathcal{D}$ is the diameter of the telescope and $\lambda=c/\nu$
is the wavelength. The Fresnel number is
	\begin{equation}
		n_F = a_F/d
	\end{equation}
where $d$ is the distance to the source.  Sources which are within
a few Fresnel scales or with Fresnel number greater than one can
be considered as nearby (technically, ``near-field").  In
Figure~\ref{fig:FresnelResponse} we plot the beam  response of an
idealized radio telescope. From this we see that sources with
$n_F>1.5$ could be detected in adjacent beams (assuming tightly
packed beams).

For the Parkes telescope, $a_F\approx 5,$km at $\lambda=21$\,cm.
In contrast, for the 305-m Arecibo dish, $a_F\approx 110\,$km. Since
Arecibo is a transit telescope sources are only found at high
elevation angles. Thus, the detection of FRB\,121102 in a single
beam at Arecibo implies that FRBs cannot be located in the troposphere
nor the stratosphere and perhaps even the mesosphere.\footnote{ The
structure of the atmosphere, starting at sea level, is as follows:
the troposphere (edge at 12\,km); the stratosphere (45\,km); the
mesosphere (85\,km); the ionosphere (1000\,km).} Given the torturous
history of FRBs it is not unreasonable to consider the possibility
that FRBs may arise from some domestic or otherwise utensil on the
International Space Station.\footnote{Near circular orbit with height of 410\,km and
inclination of $51^\circ.65$.} However, the very low declination
($-75^\circ$) of the very first event, FRB\,010724 rules out this
proposal.

The next possibility is that FRB\,121102 arises in the solar
system\footnote{This suggestion is motivated by a paper by
\citet{Karbelkar2014}. It is worth noting that the ecliptic latitude
of FRB\,121102 is $9^\circ.8$.}. However, the authors lack the
imagination and knowledge to consider this possibility.  So going
forward, we will only consider the following three possibilities
for the origin of FRB\,121102:
\begin{enumerate} 
\item[(i.)] FRB\,121102 (and by implication the Parkes FRBs) is an
ionospheric event.
\item[(ii.)] FRB\,121102 is Galactic event.\footnote{We allow for
this possibility since the spectrum of FRB\,121102 is peculiar when
compared to that of the Parkes FRBs.}
\item[(iii.)] FRB\,121102 (and by implication the Parkes FRBs) are
extra-galactic events.  
\end{enumerate}

\begin{figure}[htbp] 
   \centering
   \includegraphics[width=7.5cm]{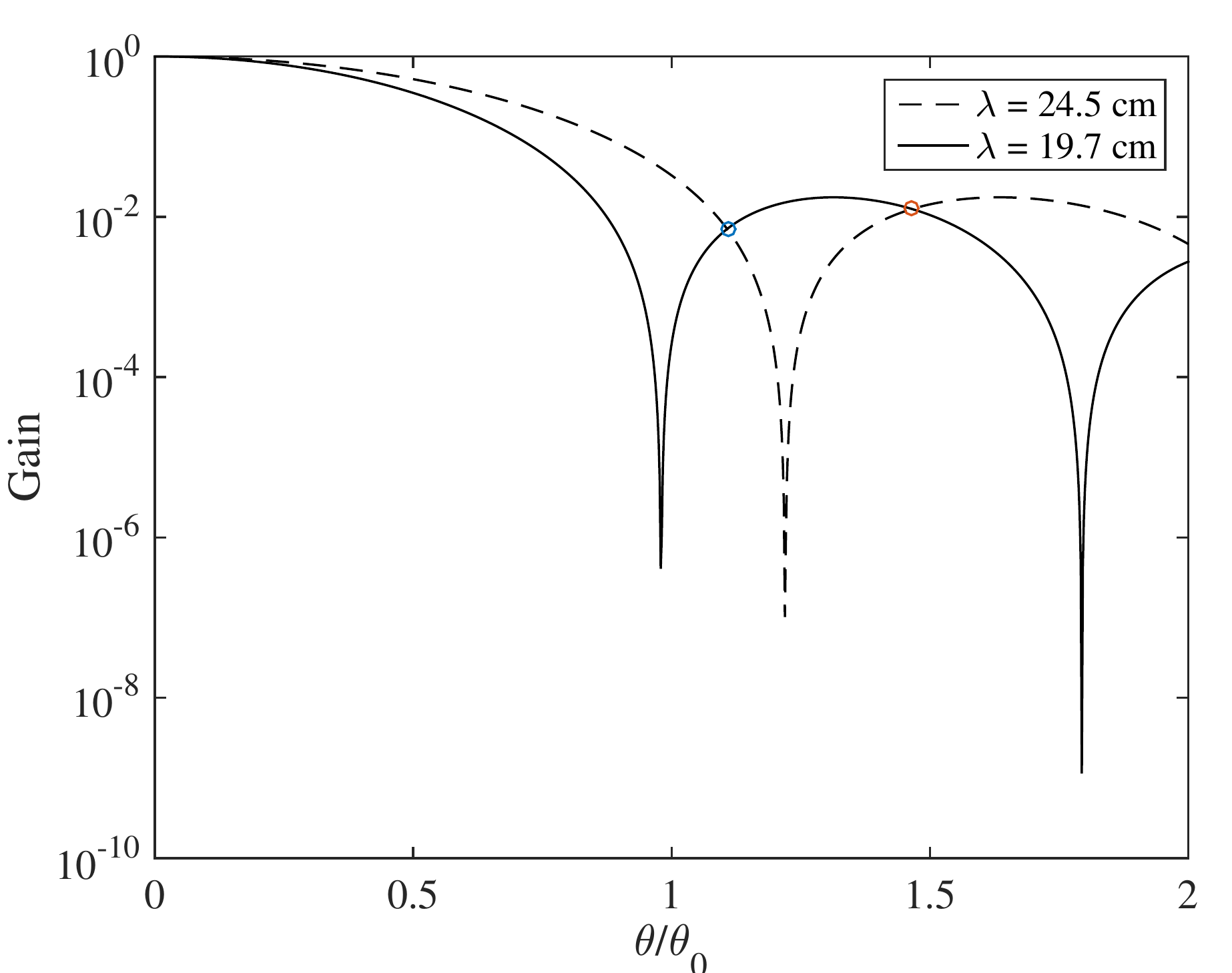}
   \caption{\small The beam response for a point source with spectral
   index, $\alpha=0$, for an unobstructed feed as a function of
   $\theta/\theta_D$ where $\theta_0=\lambda_0/\mathcal{D}$ with
   $\lambda_0=21\,$cm; as in the text, $\mathcal{D}$ is the diameter
   of an unobstructed circular aperture. The beam response is shown
   for $\lambda=19.7\,$cm and $\lambda=24.5\,$cm, corresponding to
   the two edges of the Arecibo multi-beam bandpass, 1225--1525\,MHz.
   In the angular region  bracketed by ``o" the spectral index is
   positive.  The apparent spectral index $\beta>7$ in the sky
   region in the thin annuli bounded by the radii: $[1.469, \
   1.483]\theta_0$.
	}
\label{fig:ClearAperture} 
\end{figure}

\section{The Amazing Spectral Index}
\label{sec:AmazingSpectralIndex}

There are three possible ways by which a source with an intrinsic
spectral index of, say, $\alpha \approx -1$ (such a spectral index
is routinely seen in high brightness sources), can manifest  to the
observer as a source with a  large positive spectral index: (1) due
to chromatic response of the side-lobe(s) of a telescope (2) due
to multi-path propagation in the ISM which could result in chromatic
scintillation and (3) due to  free-free absorption caused by an
intervening nebula.  Below we discuss these three options.

\subsection{Induced by the Telescope?}

Spitler and co-authors attribute the amazing spectral index to the
chromatic response of the side lobes of radio telescopes.  Recall
that the beam response of a diffraction limited telescope is a
function of $\theta/\theta_D$ where $\theta$ is the angle between
the principal axis and the line-of-sight to the source and
$\theta_D=\lambda/\mathcal{D}$ is the  diffraction angular scale.
Owing to this wavelength (frequency) dependence on the response,
as can be seen from 
Figure~\ref{fig:ClearAperture}, there are regions  in the sky where
the first side-lobe of a higher frequency band will have a larger
response relative to the first side-lobe of the lower frequency.
However, these regions are so small that the right phrase is
``apparent large spectral index is manifested in small slivers of
the sky".

\begin{figure}[htbp] 
   \centering
   \includegraphics[width=7.5cm]{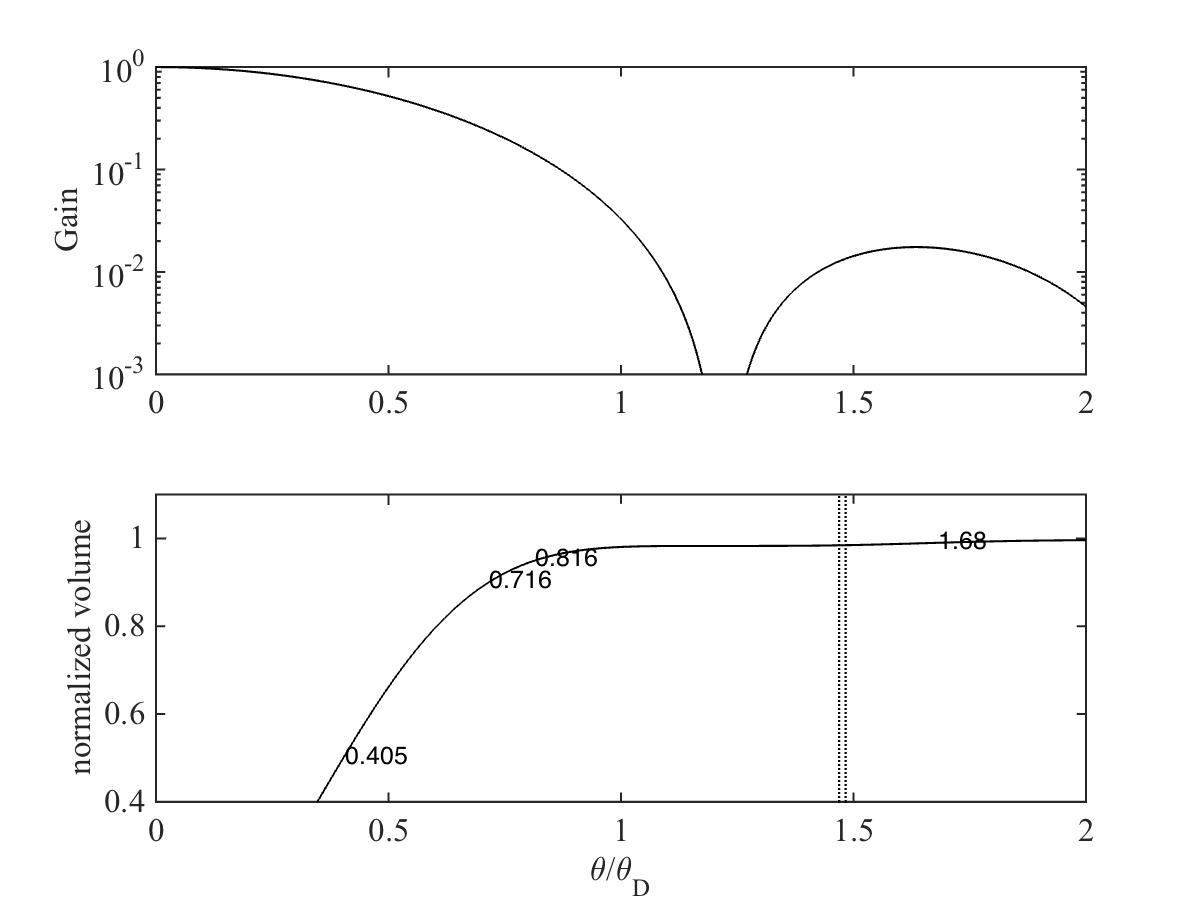} 
   \caption{\small 
   For the circular clear aperture discussed in
   Figure~\ref{fig:ClearAperture} the beam response (top) and the
   cumulative  normalized volume, $V$, as a function of $\theta/\theta_D$.
   The angular radius at which [50\%, 90\%, 95\%, 99\%] of the
   cumulative volume (normalized to unity for large angular radius)
   is displayed along the curve (from left to right). Note most of
   the sources will be found inside of the first null. The two
   vertical lines mark the angular region in which the spectral
   index exceeds 7 (see caption to Figure~\ref{fig:ClearAperture}).
   The fractional volume for $\beta>7$ is $4\times 10^{-3}$.
   }
\label{fig:EuclidianVolume} 
\end{figure}

The correct way to evaluate the fraction of sources that would
acquire a positive spectral index is to evaluate the  volume probed
by each elementary solid angle, $d\Omega$, of the Arecibo multi-beam
and then take the ratio of the sum of the volumes of those elementary
solid angles for which the response induces a large positive spectral
index to that of the total volume probed.  The elementary volume,
$dV\propto d\Omega R^3$ where $R$ is the distance to which a source
with the minimum SNR can be detected. The conversion between $S$
and $R$ requires a knowledge of the FRB population.  Let the daily
all-sky rate of FRBs with fluence greater than a given value be
represented by the following model: $N(>\mathcal{F})\propto
\mathcal{F}^q$.  For a non-evolving source population and Euclidian
geometry, $q=-3/2$.  On axis, the telescope gain is large and the
minimum SNR source has large $R$ and a correspondingly large $dV$.
For sources which are detected via side-lobes, the antenna gain is
small and this leads to a correspondingly smaller $R$ and thence
smaller $dV$ (see Figure~\ref{fig:EuclidianVolume} for a graphical
demonstration of this point).  Sources which exhibit an apparent
spectral index $>7$ (owing to chromatic beam response; see
Figure~\ref{fig:ClearAperture}) occupy a tiny fraction of the total
volume (Figure~\ref{fig:EuclidianVolume}).

\begin{figure}[htbp] 
   \centering
   \includegraphics[width=7.5cm]{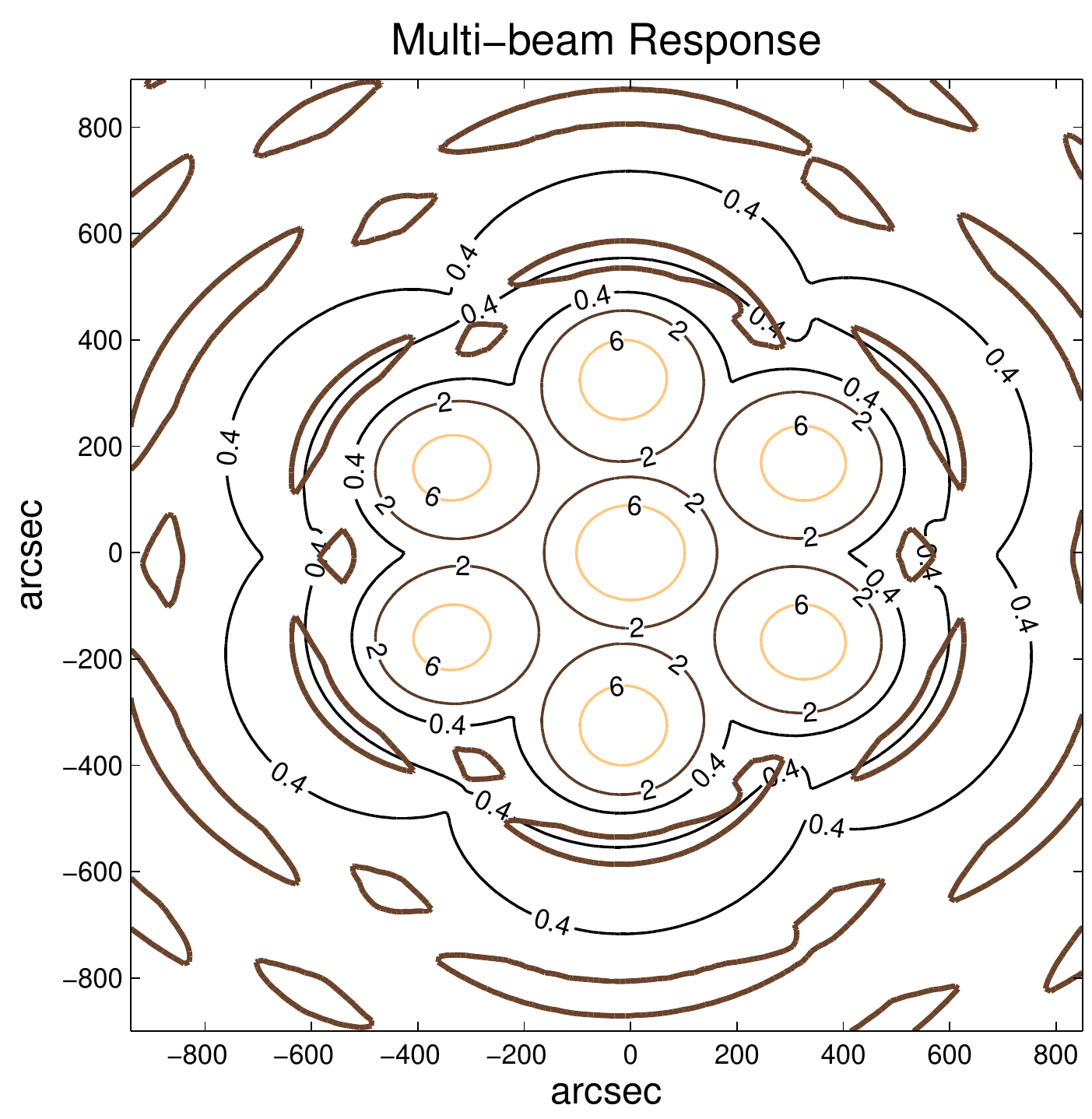} 
   \caption{\small
   The gain of the Arecibo 7-beam receiver as a function of zenith
   distance and azimuth (both in arc seconds). Only three contours
   of the gain are shown: 0.4\,K.Jy, 2\,K/Jy  and 6\,K/Jy. The
   regions of the sky where spectral indices of $7$ (or larger),
   assuming an input spectrum with spectral index of zero, can be
   produced are enclosed by dark lines. The central beam is ``beam
   0" and the six outer beams are beams 1 through 6 (clockwise).
   Since the telescope is alt-az  the sky orientation of the beams
   will depend on the hour angle (and local sidereal time). The
   beam model used here was provided by L. Spitler.
	}
\label{fig:PalfaBeam} 
\end{figure}

For a source population  with $q=-3/2$, using the accepted Arecibo
multi-beam response model, we find the following fractions: 0.4\%
and 0.1\% for $\beta=[7,11]$; see Figure~\ref{fig:PalfaBeam} for
graphical summary.  These are upper limits since for $7<\beta<11$
the band-integrated fluence for a side-lobe detection is further
reduced due to chromatic beam response (the reduction is relative
to that of an FRB which is detected in the main beam). The additional
factor of reduction is about 2 (see \S\ref{sec:BolometricFluence}
for explanation.)  The resulting additional suppression in the volume
is then $2^{-3/2}\approx 0.36$.  Thus, the odds of detecting a normal
FRB in a side-lobe (so that a large spectral index, $\beta\ge 7$
is obtained) is 1 out of 700. These odds are low enough that we
consider it is not  likely that chromatic response from the telescope
resulted in the observed large spectral index.\footnote{A simple
way to test this idea is to determine the fraction of pulsars or
RRATs with apparently large $\beta$ (relative to the entire sample)
discovered by Arecibo multi-beam survey.}

\subsection{Induced by Multi-path Propagation?}

Turbulence (density variations) in the ISM result in multiple rays
from the source reaching the observer.  This phenomenon is well
known to pulsar astronomers as ``interstellar scattering and
scintillation" (ISS).  There are two simple consequences: angular
broadening of the source and the temporal broadening of a narrow
pulse of radiation (``scattering tails").  If the turbulence is
strong then ISS can lead to frequency dependent scintillation.

According to \cite{bcc+04}, $\tau_d$, the broadening timescale at
1.4\,GHz  for a DM of 188 cm$^{-3}$\,pc ranges from $2.7\,\mu$s to
2.7\,ms.  The de-correlation bandwidth is $\Delta\nu_d =
(2\pi\tau_d)^{-1}$ and is no larger than 6\,MHz -- far too small
to account for the the doubling of frequency over more than 100
MHz.  It is quite reasonable for us to conclude that FRB\,121102
cannot be an extra-galactic source which suffered multi-path
propagation and as a result exhibited  an apparent $\beta$ of 7
to 11.

\begin{figure}[htbp] 
   \centering
      \includegraphics[width=7.5cm]{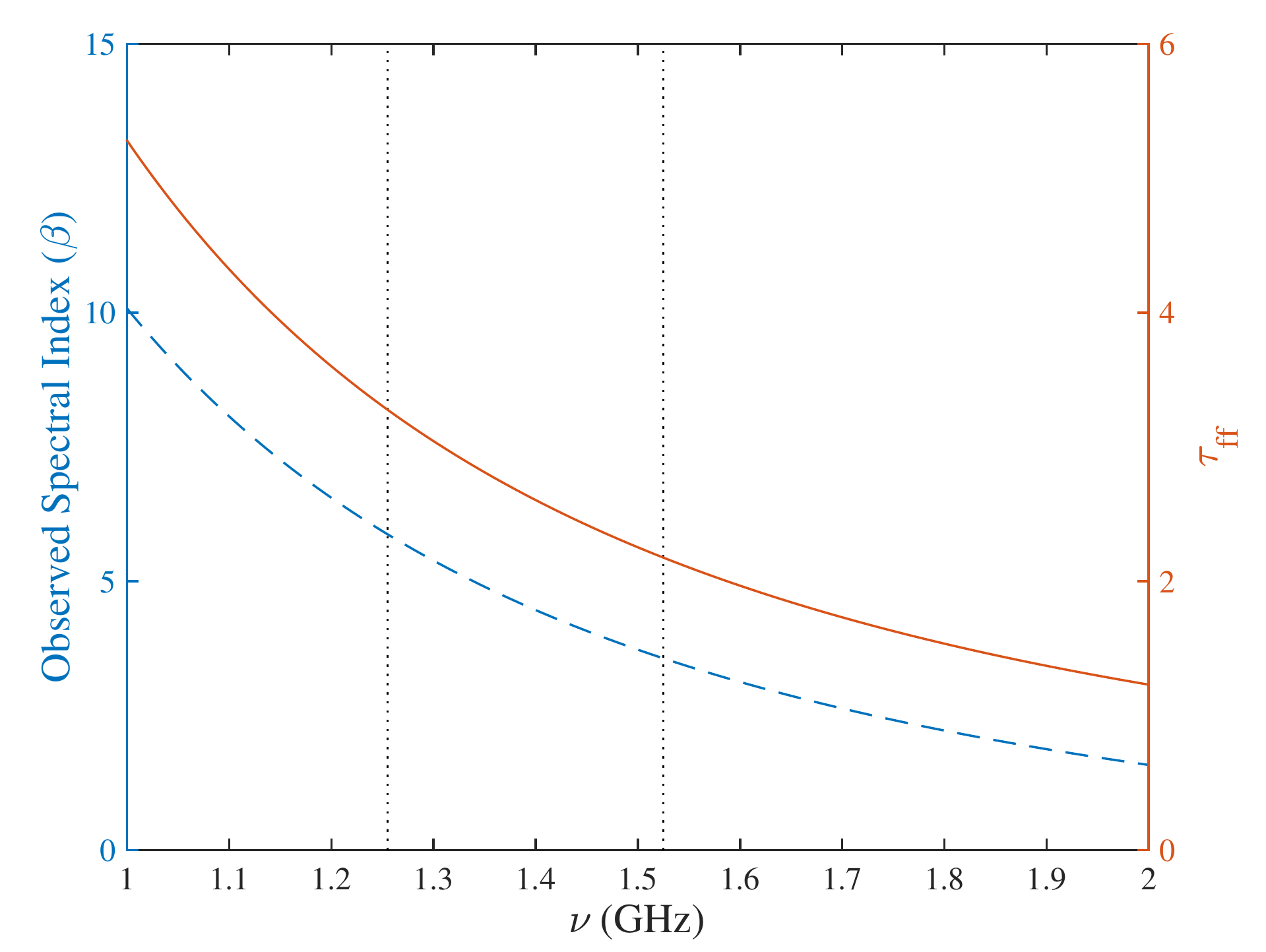}
   \caption{\small (Right y-axis; solid line): The free-free optical
   depth as a function of frequency, $\tau_{\rm
   ff}=\tau(\nu_0)(\nu/\nu_0)^{-2.1}$ with $\tau(\nu_0)$ set to
   2.5.    (Left y-axis; dashed line): The observed spectral index,
   $\beta(\nu)$ for an input power law spectrum with $\alpha=-1$;
   see Equation~\ref{eq:beta} for the definition of $\beta$.  The
   dotted vertical lines indicate the edges of the Arecibo multi-beam
   bandpass.
	}
\label{fig:ObservedSlope} \end{figure}

\subsection{Induced by Intervening Absorber?}
\label{sec:InterveningAbsorber}

In the decimetric band the primary absorption mechanism is free-free
absorption.  The free-free optical depth, $\tau_{\rm ff}$, is
proportional to $\int\phi n_e^2dl$ where $n_e$ is the density of
electrons, $\phi$ is the filling factor of clumps of electrons and
the integral is along the line-of-sight.  The constant of proportionality
is temperature dependent.  There are two possible temperature
regimes: a low temperature regime (corresponding to a photo-ionized
nebulae) and a high temperature regime (corresponding to stellar
corona).  \citet{lsm14} have proposed a model in which FRBs arise
in the corona of stars.  The combination of high temperature and 
large radius that is needed for this model will result in violent
outflows and strong X-ray emission (see \S7 of K14). 
The expected X-ray emission scales as ${\rm DM}^6$. With
FRBs now being reported with DM approaching 2,000\,cm$^{-3}$\,pc
the model is severely challenged, on  observational grounds.  
So we will not discuss it anymore
in this paper.  Here, we  consider  the low temperature option --
an intervening Galactic photo-ionized nebula.

The temperature of photo-ionized nebulae lie in a narrow range around
$T_e\sim 8,000\,$K.  In this frequency range the free-free absorption
depth is given by
	\begin{equation} 
	\tau_{\rm ff}(\nu) =4.4\times 10^{-7}{\rm
	EM}\Big(\frac{T_e}{\rm 8,000\,K}\Big)^{-1.35} \Big(\frac{\nu}{\rm
	1\,GHz}\Big)^{-2.1} 
	\label{eq:tauff}
	\end{equation}
where EM$=\int n_e^2dl$  is the ``emission measure"; the
unit is cm$^{-6}$\,pc (Lang 1980, p.\,47)\nocite{Lang80}

The spectrum of an FRB with foreground free-free absorption is then
$f(\nu) = S(\nu)\exp(-\tau_{\rm ff})$ where $S(\nu)$ is the intrinsic
spectrum.  At each frequency one can define an apparent spectral index
	\begin{eqnarray} 
	\beta(\nu) &\equiv& \frac{d{\rm log}f(\nu)}{d{\rm log}\nu}\cr
	&=& \alpha - \tau_0\big(\frac{\nu}{\nu_0}\big)^{-2.1}  
	\label{eq:beta}
	\end{eqnarray}
where the input spectrum is a power law, $S(\nu)\propto \nu^\alpha$.
In Figure~\ref{fig:ObservedSlope} we plot $\beta(\nu)$ across the
Arecibo bandpass. We choose the normalization, $\tau(\nu_0)=2.5$
so that the mean spectral index over the Arecibo bandpass is 5.  We
made this conservative choice\footnote{As can be seen 
from Equation~\ref{eq:beta}, over the range expected for $\alpha$ of say
1 to $-2$,
the inferred value of $\tau_0$ is quite robust.} since the apparent spectral
index is known to be  covariant with the
peak flux (S14).

\begin{figure}[htbp] 
   \centering
   \includegraphics[width=7.5cm]{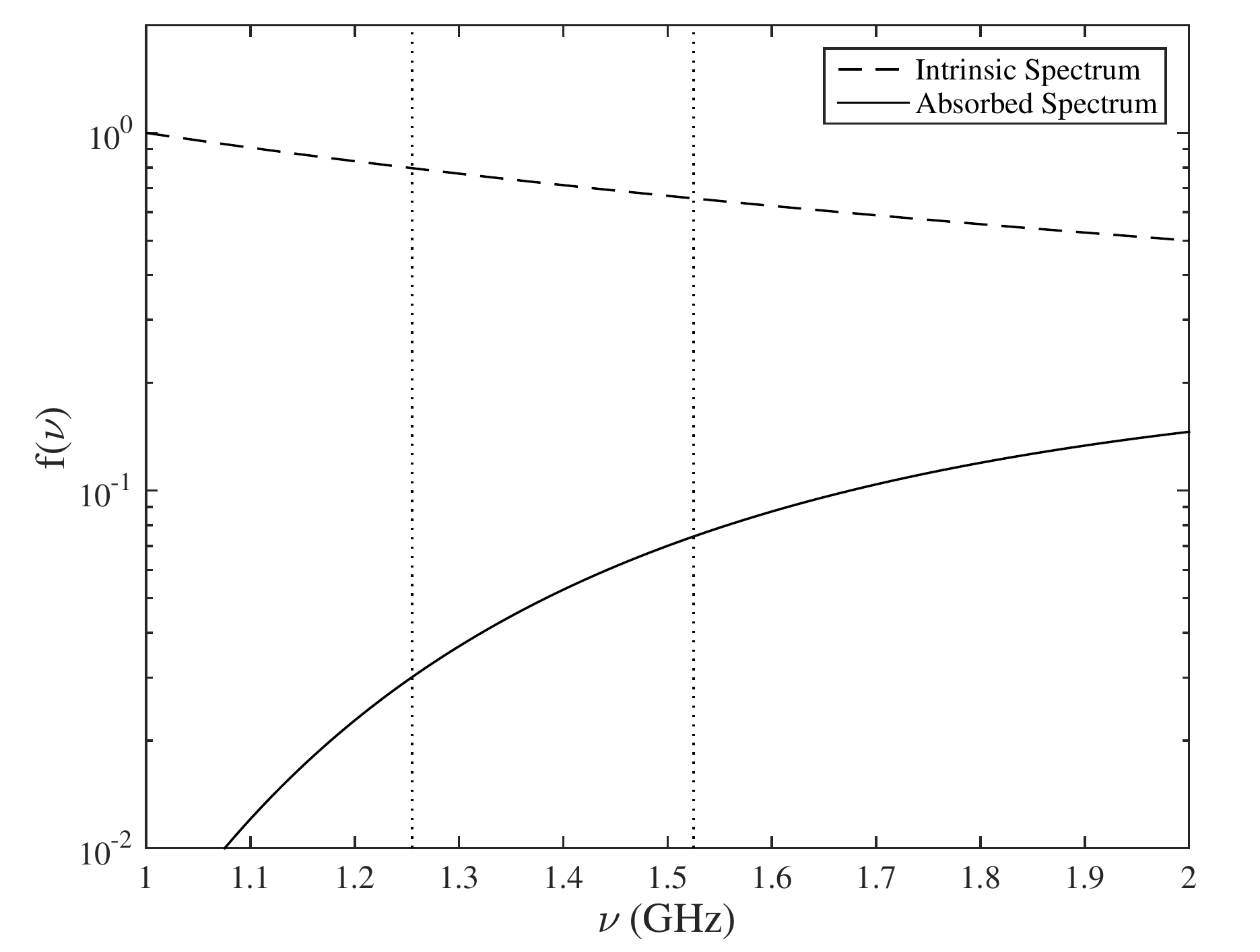}
   \caption{\small The broad-band input spectrum (dashed line) and
   the output spectrum after undergoing free-free absorption of an
   FRB with intrinsic spectral index of $\alpha=-1$. The free-free
   optical depth is $\tau(\nu_0)=2.5$.  The thin dotted vertical
   lines mark the lower and upper frequency limits of the Arecibo
   multi-beam system
	}
\label{fig:FreeFreeModel} 
\end{figure}

With the chosen normalization, using Equation~\ref{eq:tauff} we
find ${\rm EM\approx 1.2\times 10^7\,cm^{-2}\,pc}$. The resulting
absorbed broad-band spectrum is displayed in
Figure~\ref{fig:FreeFreeModel}. Note that even at the high frequency
edge the input flux is absorbed by nearly an order of magnitude.

\subsection{Summing up: Spectral Index}

In this section we considered three possible mechanisms to transmute
an intrinsic spectral index  of $\alpha=-1$ to an apparent spectral
index between 7 and 11. We reject the proposal that Galactic
interstellar medium could, as a result of ISS, produce such a large
spectral index.  We find the odds of  FRB\,121102 to be in a side-lobe
to be  1 in 700 (relative to an FRB found in the main beam) and
thus strongly disfavor chromatic side-lobe response as being responsible for
the large observed spectral index. By process of elimination we
arrive at the hypothesis that the positive spectral index is due
to absorption by an intervening nebula.  There are two possibilities
for the location of the intervening nebula: Galactic or extra-galactic.
In the next section we investigate the properties of the putative
Galactic nebula.

\section{The Intervening Nebula (Galactic)}
\label{sec:InterveningNebula}

In this section, accepting the framework in which FRB\,121102
suffered from significant free-free absorption, we now attempt to
infer the physical properties of the intervening free-free absorbing
nebula.  In part this hypothesis is motivated by the fact  that
FRB\,121102 is located at low Galactic latitude, admittedly in the
Galactic anti-center.

The primary parameters  for the intervening Galactic nebula model
are the following: $L$,  the thickness of the nebula (the length
along the line-of-sight); $D$, the diameter in the plane of the
sky; and $d$, the distance from us to the source. Our challenge is
to see if observations can constrain these parameters (and better
still if the nebula is actually detected in archival data  or could
be detectable by near-future observations).  We make the reasonable
assumption that $L$ is comparable to $D$ (spherical approximation).
Sheets are not uncommon structures in the ISM and so $D\gg L$.
However, on purely probabilistic ground, $D\ll L$ is not likely.

The two constraints for the putative nebula are the free-free optical
depth (${\rm \tau_{ff}\propto EM}$; see Equation~\ref{eq:tauff})
induced by the nebula, and the dispersion measure (DM$^\prime$)
arising inside the nebula. For the former we have already determined
${\rm EM\approx1.2\times 10^7\,cm^{-6}\,pc}$
(\S\ref{sec:InterveningAbsorber}). There is little reason to believe
that there will be significant contribution to EM by the diffuse
Galactic ISM.  The inferred DM of $557\,{\rm cm^{-3}\,pc}$ is due
to (i) Galactic contribution and (2) contribution from the nebula.
We assign ${\rm DM}^\prime=400\,{\rm cm^{-3}\,pc}$.

Both DM$^\prime$ and EM are line integrals of moments of electron
distribution along the path to the source.  We derive two fundamental
parameters of the nebula:
	\begin{eqnarray}
	n_e &=& \frac{\rm EM}{\rm DM^\prime} = 3\times 10^4\,{\rm cm^{-3}},
	\label{eq:ne}\\
	\phi L &=& \frac{\rm DM^{\prime 2}}{\rm EM} = 0.013\,{\rm pc}.
	\label{eq:L}
	\end{eqnarray}
	Going forward we will set $\phi=1$. 
The electron density is low enough that the  plasma frequency,
$\nu_p=\sqrt{n_e e^2/(\pi m_e)}=1.6\,$MHz, is well below the observing
frequency.  The mass in the nebula (assuming only Hydrogen) is
 $\approx  10^{-3}(D/L)^2\,M_\odot$.

Since the nebula is optically thick in the band under consideration
($\tau\approx 2.5$; see \S\ref{sec:InterveningAbsorber}),
the surface brightness is simply given by the Rayleigh Jeans formula:
	\begin{equation}
	I(1.4\,{\rm GHz}) = 12 (T_e/8,000\,{\rm K)\,mJy\,arcsec^{-2}}.
	\label{eq:RayleighJeans}
	\end{equation}
In the H$\alpha$ line the surface brightness is related to EM as
follows (see K14)
	\begin{equation}
		I({\rm H\alpha}) = 1.09\times 10^{-7} {\rm EM\ erg\,cm^{-2}\,s^{-1}\,sr^{-1}}.
	\end{equation}
Given that ${\rm EM\approx 1.2\times 10^7\,cm^{-6}\,pc}$
(\S\ref{sec:InterveningAbsorber}) we find the expected surface brightness is
	\begin{equation}
	I({\rm H\alpha}) = 3.1\times 10^{-11}\,{\rm erg\,cm^{-2}\,^{-1}\,arcsec^{-2}}.
	\end{equation}
It is illustrative to convert this intensity to photon units: $I({\rm
H\alpha}) \approx {\rm 10.1\  photon\ cm^{-2}\,s^{-1}\,arcsec^{-2}}$.
In contrast, the night sky (new moon) has a brightness of
21\,mag\,arcsec$^{-2}$ in the R-band. This corresponds to a photon
surface brightness of $4\times 10^{-3}\ {\rm photon\ cm^{-2}\,s^{-1}}$.

The angular size of the nebula is 
	\begin{equation}
	\theta_N = 2.75^{\prime\prime}(D/L)d_{\rm kpc}^{-1}
	\label{eq:theta}
	\end{equation}	
where $d_{\rm kpc}$ is the distance to the nebula in pc.
Integrating $I({\rm H\alpha}$) over the angular extent of the nebula
yields an H$\alpha$ flux density of
	\begin{eqnarray}
		f({\rm H\alpha}) &=& I({\rm H\alpha})(\pi/4)\theta_N^2\cr
		&=&1.83\times 10^{-10}(D/L)^2 d_{\rm kpc}^{-2}\,{\rm erg\,cm^{-2}\,s^{-1}}.
		\label{eq:fHalpha}
	\end{eqnarray}
In the 21-cm band, by construction, the nebula is optically thick 
and so the Rayleigh-Jeans formula (Equation~\ref{eq:RayleighJeans})
provides the expected surface brightness. The flux density is then
	\begin{eqnarray}
	S &=& \frac{\pi}{4}\theta_N^2 I(1.4\,{\rm GHz})\cr
	&=& 71\,(T_{\rm e}/8,000\,{\rm K})(D/L)^2 d_{\rm kpc}^{-2}\,{\rm mJy}
	\label{eq:SFlux}
	\end{eqnarray}

The number of recombinations per second is 
	\begin{equation}
	\dot N_R \approx\frac{\pi}{4} \alpha_B D^2 L n_e^2 = 0.54\times 10^{46} (D/L)^2\,{\rm s^{-1}}.
	\end{equation}
Here, $\alpha_B=1.1\times 10^{-13}(T_e/8,000{\rm K})^{-0.89}\,{\rm
cm^3\,s^{-1}}$ is the Case B recombination \citep{of06}.  If the
nebula was photo-ionized then the corresponding {\it minimum}
ionizing (UV) luminosity is
	\begin{equation}
	L_{UV}>4\times 10^{35}\,(D/L)^2\,{\rm erg\,s^{-1}}.
	\end{equation}
If a similar continuum flux is available at say 2500\,\AA\ and
4500\,\AA\ then the AB magnitude will be $13.7 + 5\log(d_{\rm kpc})$
and $13.0+5\log(d_{\rm kpc})$  mag, respectively,


\subsection{Search for Counterpart}

Next, we conducted a search of relevant archival data sets for the
counterpart of the putative nebula. Specifically we restricted the
search to a circular region of radius 413 arc seconds and centered
on beam 4. The radius is large enough to include the side-lobe
response to 0.4\,K/Jy (which incidentally is approximately equal
the on-axis gain of the Parkes telescope; see Figure~\ref{fig:PalfaBeam}).
Below we describe the result of this search.

\begin{deluxetable}{lllrr}
\tabletypesize{\footnotesize}
\tablecolumns{5}
\tablewidth{0pt}
\tablecaption{NVSS Sources in the field of FRB\,121102
\label{tab:Sources}}
\tablehead{\colhead{Name} &
\colhead{RA} & \colhead{Dec} & \colhead{$\Delta\theta$} & \colhead{Flux (mJy)} }
\startdata
AFRB & 05 32 9.6 & +33 05 14 & 0$^{\prime\prime}$        & --\\
VLA1 & 05 32 09.7 & +33 04 05 &   $69^{\prime\prime}$     & $3.2\pm 0.4$\\
VLA2 & 05 31 53.8 & +33 10 15 &   $308^{\prime\prime}$   & $4.6\pm 0.5$\\
\enddata
\tablecomments{The nominal position of FRB\,121102 (``AFRB"). The
search radius was 413 arc seconds. The two sources, VLA1 and VLA2
are from the NRAO VLA Sky Survey (NVSS). $\Delta\theta$ is the
radial offsets of the VLA sources from AFRB. 
	} 
\end{deluxetable}

\subsubsection{Decimetric Band}

The fruits of our search of the VLA Sky Survey \citep{ccg+98} -- a
survey which mapped the entire Northern Sky in the 21-cm band --
can be found in Table~\ref{tab:Sources}.
The two sources are nominally noted to be unresolved in the NVSS
catalog and do not correspond to any source in the IPHAS catalog (see \ref{sec:IPHAS}).  Likely they
are background radio sources. Nonetheless, comparing
Equation~\ref{eq:SFlux} with $D\approx L$ to the flux of the two
sources we find $d_{\rm kpc}>4.2\,$kpc.

\subsubsection{IPHAS}
\label{sec:IPHAS}

Using the Isaac Newton Telescope and a mosaic of CCDs the  Northern
Galactic plane was imaged -- the Isaac Newton Telescope Photometric
Halpha Survey or IPHAS \citep{dgi+05}. The survey was undertaken
in three bands: $r^\prime$, $i^\prime$ and a custom narrow-band
filter centered on H$\alpha$ at zero velocity (6563\,\AA) and a
full-width-at-half-maximum (FWHM) bandpass of 95\,\AA\ (hereafter, we will
refer to this band as {\it ha}). All the magnitudes are referenced
to the Vega system and the assumed zero point(s) are given in
Appendix \ref{sec:Halpha}.

We inspected the {\it ha} image carefully and found no evidence of
an extended nebula, as would be expected for a source located at
say 1\,kpc (cf. see Equation~\ref{eq:theta}). This motivates us to
consider a source at, say, 10\,kpc. In this case, the nebula will
appear point like.

\begin{figure}[htbp] 
   \centering
   \includegraphics[width=7.5cm]{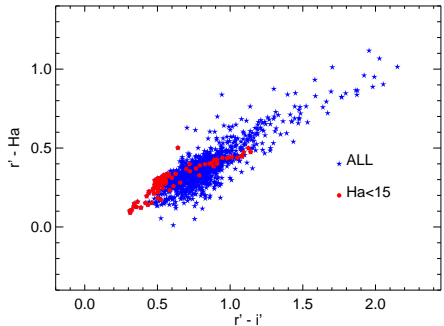} 
   \caption{\small 
   The $r^\prime - $H$\alpha$ ({\it ha}) and $r^\prime-i^\prime$ color-color diagram for
   the IPHAS sources in the error circle of FRB\,121102. We examined
   all objects with {\it ha}\ $ <15.0$ (shown as red filled circles).
   They are consistent with being ordinary stars.
      }
\label{fig:CMD} 
\end{figure}

Combining Equation~\ref{eq:fHalpha} and Equation~\ref{eq:HalphaMag}
we obtain the following equation:
	\begin{equation}
	{\it ha} = 7.8 +5\log(d_{\rm kpc})
	\end{equation}
Thus, even for a source at 20\,kpc (a plausible distance that places
the source at the known edge of the Galaxy) we would expect to see
bright H$\alpha$ source, ${\it ha}\approx 14.3$. To be conservative,
we examined all objects with {\it ha} brighter than ${\it ha} = 15$
(red filled circles in Figure~\ref{fig:CMD}).

This figure is identical to Figure 1 in \cite{wkg+06}, in which the
authors use IPHAS photometry to identify H$\alpha$ emitting sources.
Their figure shows that normal stars with magnitudes brighter than
$r^\prime = 16$ occupy a small region along the line delineated by the majority
of objects, while outliers above this cloud are candidate H$\alpha$
emitters.  The scatter in the stellar distribution is due to variations in
the extinction along the line-of-site to stars of varying distances.
Figure \ref{fig:CMD} shows only one candidate outlier above the
distribution shown by the red filled circles, which is
TYC 2407-679-1, an A star (B$-$V$\cong 0.3$\,mag).

We also searched the catalog for objects that were detected in $ha$,
but not in the other filters and only two objects met this criterion,
J053232.67+330840.0 at $ha = 19.3$, and J053211.51+330419.8 at $ha
= 19.4$.  Both of these objects are too faint to be the nebular
counterpart that produced the FRB.

\begin{figure*}[htbp] 
   \centering
   \includegraphics[width=6in]{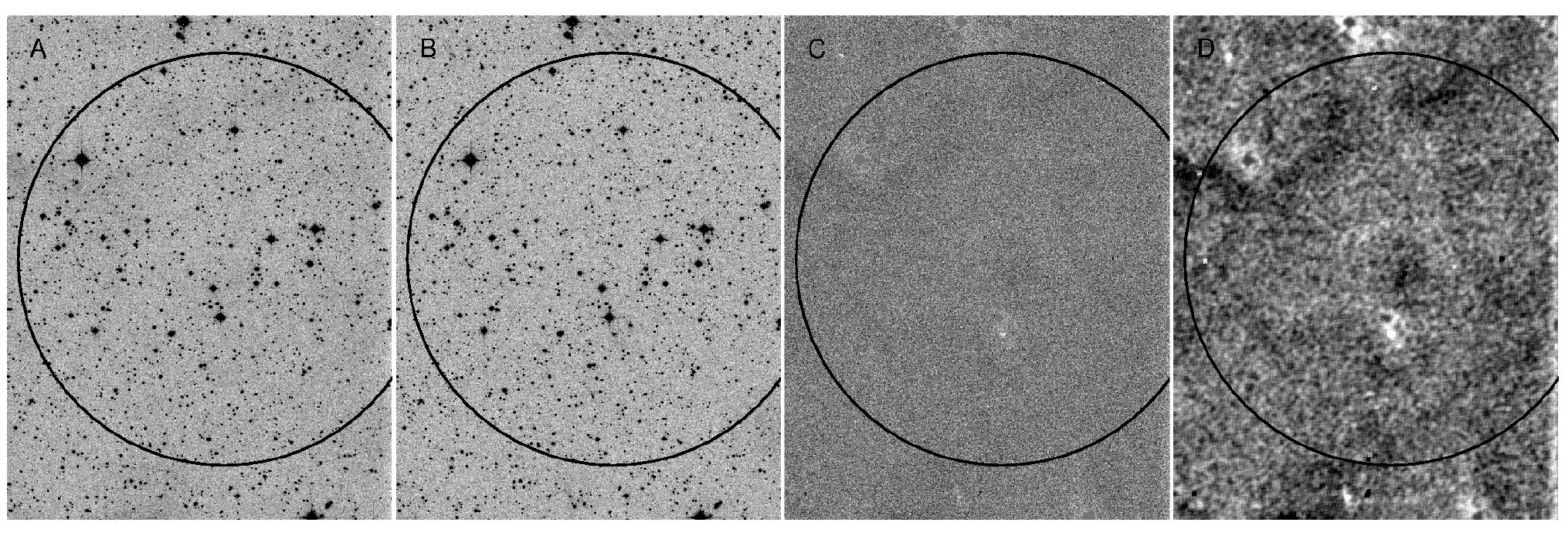}
   \caption{\small
   From left to right.  
   [A] Co-addition of 14 PTF H$\alpha$ (656\,nm)
   images.  
   [B] Co-addtion of 14 PTF H$\alpha$-off (663\,nm) images.
   [C] The on-off subtracted image using a custom image subtraction
	algorithm (Zackay, Ofek \&\ Gal-Yam, in prep).
   [D] The subtracted image smoothed by a square of side
    $10^{\prime\prime}$ top hat filter to identify faint-extended
   features.  The grayscale is inverted. In particular, for panel C and D,
   positive signal is black. The white patches are artificats due to
   wings of bright stars.
   }
\label{fig:AFRBPTF} 
\end{figure*}

\subsubsection{The PTF H$\alpha$ Survey}

The primary goal of the Palomar Transient Factory \citep{lkd+09,rkl+09}
is a systematic study of the transient optical sky. To this end the
dark and gray time is used to undertake a number of ``experiments"
in $g$ and $R$ bands. 
However, during three nights around full moon, the PTF collaboration has been
undertaking a Census of the Local Universe (CLU) survey to complete our
catalog of nearby galaxies. CLU deploys four contiguous narrow-band filters
to search for extended, H$\alpha$ emitters across 3$\pi$ of the Northern
sky out to a distance of 200\,Mpc \citep{k11}. CLU assists the discovery of
local Universe transients by providing a host galaxy redshift which is a
severe filter to channel follow-up. For example, CLU helps pinpoint
electromagnetic counterparts to gravitational waves in two ways (i) for
wide-angle optical searches, CLU promptly reduces false positives by two
orders of magnitude \citep{kn14,nkg13} (ii) for narrow-field X-ray/infrared
searches, CLU helps determine the top pointings \citep{gck+15}. 

For the purpose of the discussion here only two bands are relevant:
centered on the rest frame H$\alpha$ at 656\,nm (``H$\alpha$-on")
and the other on 663\,nm (``H$\alpha$-off").  PTF obtained 14  images
in each one of the filters.  We co-added the images taken in each
filter using {\tt SWarp} \citep{bertin2010}.  The images are presented
in Figure~\ref{fig:AFRBPTF}.

The basic data reduction of PTF data  is described in \citet{lsg+14}.
The PTF H$\alpha$ survey is still being calibrated. We undertook
the following {\it rough} calibration.  Following the prescription
in \citet{oll+12} we calibrated the image using the Tycho 2 star
at RA=$83^\circ.07877$ Dec=$+33^\circ.192993$.  Fitting its $B_{T}$,
$V_{T}$, $J$, $H$, $K$ magnitudes with the \citet{Pickles98} main
sequence stars spectral templates, we found a best fit with a
spectral class F5V.  Using its spectral template we calculate the
star H$\alpha$ flux from synthetic photometry of an F5V template.
We found the star AB magnitudes are 11.87 and 11.80 in H$\alpha$-on
and H$\alpha$-off bands, respectively.   The Zero Point (ZP) for
the H$\alpha$ image is 24.05 AB mag.

A number  of point sources exhibit H$\alpha$ in emission.  The point
sources range from 17.3 AB mag to 19.3 AB mag (the detection limit).
In an effort to search for diffuse emission we smoothed the difference
image.  The brightest sources in the nebulosity in the image are
about ~20 AB mag per $10\times 10$ arcsec$^2$ (corresponding to
0.036 mJy).  In summary both the IPHAS and PTF analysis shows that
any putative intervening nebula is beyond 20\,kpc -- beyond the edge
of the Galaxy in this direction.

\subsection{Summing Up: Galactic Nebula}
 
In summary, we searched for a putative Galactic compact nebula which
could produce the necessary free-free absorption. We found no
evidence in the radio (free-free continuum) nor H$\alpha$ (recombination
radiation). We can rule out that the necessary nebula does not
exist, at least within the galaxy (say a distance of  up to 20\,kpc
which would certainly mark the edge of the Galaxy in this direction).
So we conclude that were FRB\,121102 to be extragalactic then the
intervening nebula must be located in the host galaxy of FRB\,121102.
The inferred density of the intervening nebula (see Equation~\ref{eq:ne})
is clearly atypical for diffuse ISM. FRB\,121102 must thus be
associated with a region of rich ISM (and thence high star-formation).

On a separate but important issue we note that the failure to find
a compact nebula also rules out a purely Galactic explanation for
FRB\,121102, namely, a Rotating Radio Transient (RRAT) with a
conveniently arranged intervening nebula (cf.\ the radio burst
discussed by \citealt{ksk+12}).

\section{The Locale of FRBs}
\label{sec:LocaleOfFRBs}

Here, we gather our thoughts on the origin of FRBs. By  treating
the Parkes and the Arecibo FRBs as a single class, we can state
with some confidence that FRBs must be located beyond 100\,km. This
then opens up three reasonable locales: the ionosphere, the Galaxy
and other galaxies.  In the ionosphere model, a ball of charged
particles somehow produce millisecond bursts in the decimetric band
and whose emission follows a $t_a\propto \nu^m$ with $m=-2$ to good
precision.  Furthermore, in some cases the pulse has to be broadened
by milliseconds (ms). The usual explanation for ``scattering" tails
is multi-path propagation. 
The scattering time scale of a radio pulse is given by
\begin{equation}
\tau_{{\rm d}} = \frac{1}{2c}\theta^{2}\frac{D_{l}D_{s}}{D_{ls}},
\label{eq:tau_d}
\end{equation}
where $D_{l}$, $D_{s}$ and $D_{ls}$ are the
distances to the scattering screen, the source
and scattering screen to the source, respectively.
Due to the scattering the source angular size will be
amplified to $\theta$.
An important point is that if the source is nearby, in order to get
a scattering over 1\,ms, the angular size of the source will
have to be larger than the beam size, and hence it will be detected
in multiple beams.
Therefore, we can use this to set a lower limit on the distance to FRBs.
Setting $\tau_{{\rm d}}=1$\,ms, and $\theta<11$\,arcmin,
we find that $D_{s}\gtorder10^{-3}$\,AU which far exceeds the size
of the ionosphere.
So, as with the measured DM, we need to find an alternative explanation for
the scattering tails.

We have thoroughly investigated a possible Galactic origin (in which
most of the inferred intervening electrons arise in a Galactic
nebula) for FRB\,010724 (K14) and here we have undertaken a similar
investigation for FRB\,121102. In both cases, we did not find any
evidence to support a Galactic origin.

We thus find ourselves between two exotic possibilities:  (i) an
entirely unknown ionospheric phenomenon with particularly exotic
(a more accurate word is  ``tailored") properties or (ii) an
extra-galactic phenomenon in which the frequency dependent arrival
time, scattering tails and occasional free-free absorption can be
readily accounted. The primary challenge for the extra-galactic
model is that the FRB pulses are as brilliant as those of pulsars
but last {\it six orders of magnitude} longer or equivalently arise
in regions which are six orders of magnitude larger than those of
pulsars. The energetics in the radio alone is staggering.  Two
decades ago astronomers were in a  similar quandary in regard to
the origin of GRBs: an increasingly exotic local model and a simpler
extra-galactic model but with extra-ordinarily large power and
energy demands.  We have exhausted all reasonable Galactic options
and being forced to consider ``designer" models with increasingly strident
demands for the terrestrial option. Guided by the history of our field,
we are driven to the extra-galactic model. The rest of the paper
is an exploration of the extra-galactic model.

\section{An Extra-Galactic Origin}
\label{sec:AnExtraGalacticOrigin}

In the extra-galactic model for FRB\,121102, the dispersion measure
in excess of the Galactic contribution is about ${\rm 370\,cm^{-3}\,pc}$.
This is due to contribution from electrons in the  intergalactic
medium, electrons in the diffuse interstellar medium (ISM) and from
electrons in the dense nebula close to the FRB source. The reason we
are forced to invoke a dense nebula is that neither the IGM nor the
host diffuse ISM (along most lines of sight) would have an emission
measure so high as to produce $\tau(\nu_0)\approx 2.5$.

This intervening EM-inducing nebula will also contribute to the
dispersion measure.  Lacking any additional information, we apportion
the excess dispersion measure (above) about equally to the IGM (say,
${\rm DM_{IGM}\approx 170\,cm^{-3}\,pc}$) and to the dense nebula
(DM$^{\prime\prime}\approx 200\,{\rm cm^{-3}\,pc}$).

In the extra-galactic framework, the  distance to an FRB, $d_{\rm IGM}$,
is inferred from ${\rm DM_{IGM}}$ and the use of an accepted model for
the mean density of the IGM plasma as a function of redshift (e.g.\
\citealt{ioka03,zok+14}).  With this assignment the inferred redshift
to the host galaxy is reduced to $z\approx 0.19$.  The density and
the width of the intervening plasma, relative to that discussed in
\S\ref{sec:InterveningNebula}, is a factor of two higher and a
factor of four lower respectively (see Equations~\ref{eq:ne} and
\ref{eq:L}).  Thus, $n_e=6\times 10^4\,{\rm cm^{-3}}$ and $L=3.3\times
10^{-3}\,$pc.

In the absence of free-free absorption the peak flux of FRB\,121102
is about ten times higher than that observed (see
Figure~\ref{fig:FreeFreeModel}).  FRB\,121102, had it not been for
the absorption by the compact nebula, would have been as bright as
the typical Parkes FRB.

\subsection{Volumetric Rate of FRBs}
\label{sec:VolumetricRate}

At the present time, FRBs are localized to fractions of square degrees
and this localization is simply not good enough to securely identify
FRBs with other galaxies.  Let $\mathcal{R}$ be the daily all-sky
rate of detected FRBs.  According to \citet{kp14},
	\begin{equation}
	\mathcal{R}(\mathcal{F}\ge 2\,{\rm Jy\,ms}) \approx 2500\,{\rm day}^{-1}.
	\label{eq:FRBDailyRate}
	\end{equation}
With the distances inferred from ${\rm DM_{IGM}}$ (\S\ref{sec:AnExtraGalacticOrigin}),
the inferred the volumetric rate of FRBs is
	\begin{equation}
		\Phi_{\rm FRB}(\mathcal{F}>2\,{\rm Jy\,ms})=6\times 10^3\,{\rm Gpc^{-3}\,yr^{-1}};
	\label{eq:PhiFRB}
	\end{equation}
see K14 for details.
Any proposed extra-galactic model for FRBs must demonstrate that a
radio signal can propagate from the explosive site without significant
absorption. Separately, the same model must account for the FRB
volumetric rate.

Following K14, on physical grounds we reject both major families
of supernovae (core-collapse, Type Ia) and long duration gamma-ray
bursts (both low and high luminosity).  The volumetric rate of
short-hard bursts is clearly insufficient to account for $\Phi_{\rm
FRB}$. The  {\it blitzar} model in which FRBs originate from massive
NS collapsing into BHs was  cleverly designed to avoid the problem
of significant free-free absorption \citep{fr14}.  However, this
model cannot be reconciled with the demographics of Galactic pulsars
and also that of  bright supernovae (K14). So we reject the blitzar
model as well.  In the next section we explore the model which
avoids the key physical fatal flaw (local absorption) and the
physical requirement of possessing sufficiently large volumetric
rates.

\section{Giant Flares as causing FRBs}
\label{sec:GFFRB}

There are three reasons why we find the model in which FRBs result
from Giant Flares from SGRs. First, as with pulsars there is very
little absorbing matter in the immediate vicinity of SGRs.  Thus,
there is no impediment to the transmission of radio pulses.  Next,
the volumetric rate of giant flares  is well matched to that of
FRBs (see below).  Finally, a physically plausible model in which
giant flares from SGRS give rise to FRBs has been proposed
\citep{lyubarsky14}.

According to \citet{Ofek07} the volumetric rate of giant flares from SGRs is as follows:
	\begin{eqnarray}
	\label{eq:HGF}
	\Phi_{GF}(\mathcal{E}_\gamma\lesssim \mathcal{E}_*)&<&2.5\times 10^4\,{\rm Gpc^{-3}\,yr^{-1}}
	\\
	\label{eq:GF}
	\Phi_{GF}(\mathcal{E}_\gamma\gtrsim \mathcal{E}_1)&\approx& 4\times 10^5(\tau_{\rm GF}/25\,{\rm yr})^{-1}\,{\rm Gpc^{_3}\,yr{-1}}.
	\end{eqnarray}
where $\mathcal{E}_*=3.6\times 10^{46}\,$erg, $\mathcal{E}_1=3\times
10^{44}\,$erg  and $\tau_{\rm GF}$ is the mean time between giant
flares; $\tau_{\rm GF}>30\,$yr at the 95\% confidence level
\citep{Ofek07}. We will refer to flares with $\mathcal{E}\approx
\mathcal{E}_*$ as ``hyper-giant" flares (HGF) and those with energy
comparable to $\mathcal{E}_1$ as giant flares.

The isotropic energy release of a  1\,Jy FRB with $\alpha=-1$ and
lasting 1\,ms is $10^{39}d_{\rm Gpc}^2\,$erg; here $d_{\rm Gpc}$
is the distance to the FRB in units of Gpc. The unknown and key
issue is the efficiency, $\eta$, with which the giant flare energy
(primarily in $\gamma$-rays) can be converted to a brilliant radio
pulse.  If $\eta$ can be as large as $10^{-5}$ then there is
comfortable margin between the volumetric rate of detected FRBs
(Equation~\ref{eq:PhiFRB}) and that of the giant flares from SGRs
(Equation~\ref{eq:GF}). If $\eta$ is much smaller, say $10^{-7}$
then the relevant SGR rate is that of Hyper-giant flares
(Equation~\ref{eq:HGF}) and the margin between the FRB rate and the
HGF rate is less than a factor of four.

\subsection{A Young Magnetar Model}
\label{sec:YoungMagnetarModel}

In our Galaxy giant flares have been seen from two young magnetars:
SGR\,1806$-$20 (age, $650\pm 300$\,yr) and SGR\,1900+14, age $(6\pm
1.8)\times 10^3\,$yr \citep{tck13}.  Both these magnetars are
associated with star clusters (see \citet{tck13} for summary).
A dense ISM is seen in the vicinity of SGR\,1806$-$20 and a near-IR
dust ring or shell has been noted around SGR\,1900+14 \citep{wrd+08}.
Motivated by this basic observation we propose that FRBs result
from {\it young} magnetars. A natural consequence of this model is
the presence of a rich ISM on scales of a parsec (as in SGR\,1900+14)
to a size set by region of intense star formation (tens to hundreds
of parsecs).

The young magnetar model can readily explain the presence of a dense
ISM, as inferred for the case of FRB\,121102. Most FRBs exhibit
an exponential decay with timescales, $\tau_d\sim
1\,$ms. This ``tail" is attributed to multi-path propagation induced
by turbulent structures. The young magnetar hypothesis naturally
accounts for such scattering tails.

Star-forming regions, specifically HII regions and supernovae shells
are known for having turbulent structures.  The  angular broadening,
$\theta_s$, at the fiducial frequency,  expected from the inferred
turbulence (``scattering measure") in prominent Galactic HII regions
(e.g.\ NGC\,6334; Cygnus star forming region; Galactic Center) is
$\theta_s\sim 0.1$--$1^{\prime\prime}$ (see \S11 of K14).  Angular
broadening leads to multi-path propagation with an exponential timescale
of
	\begin{equation}
		\tau_d = \frac{D_s}{2c}\theta_s^2 \label{eq:taud}
	\end{equation}
where $D_s$ is the distance from the FRB (magnetar) to the turbulent
medium.  Thus, a turbulent medium at, say, $D_s\sim$ 10--100\,pc
would easily account for the scattering tails.

\subsection{The Orion Star-forming Region}
\label{sec:Orion}

In order to concretely understand the ramifications of the young
magnetar model, some familiarity of the locales (in this hypothesis)
would be useful. To this end, consider the Orion nebula.  It is
not a particularly rich cluster (relative to those of SGR\,1900+14
and SGR\,1806$-$20). However, it is one of the best studied objects
in modern astronomy.

The diameter of Orion nebula is 8\,pc. The peak optical depth at
1.4\,GHz is  is  0.055 and the corresponding EM is $2.5\times
10^5\,{\rm cm^{-3}\,pc}$ \citep{fcw+93}. Thus, the mean density of
Orion is $177\,{\rm cm^{-3}}$ and the corresponding maximum DM
would be 1416\,cm$^{-3}$\,pc.  However, the structure of Orion
nebula is, as is the case with other star forming regions, complex.
On one side, the newly born cluster of hot stars are irradiating
and ionizing the parental molecular cloud and this in turn is likely
to initiate new star formation inside the cloud. On the opposite
side the ionized gas is flowing away from the star cluster.

The ionization structures of the Orion nebula are summarized in
Table~\ref{tab:Orion}. The magnetic field in Orion is comparable
to the thermal pressure \citep{Ferland01} or about 0.4\,mG. The
resulting EM, DM and RM are summarized in Table~\ref{tab:Orion}.
The point of this table is to expose the reader to the complexities
of real star clusters (and their rich ISM) and, in particular, to the
potentially large contribution to DM, EM and even rotation measure
(RM; see \S\ref{sec:Ramifications} for definition of RM).

\begin{deluxetable}{rrrrrr}
\tabletypesize{\scriptsize}
\tablecaption{Physical parameters for the Orion Nebula}
\tablewidth{0pt}
\tablehead{
\colhead{Zone} & \colhead{$n_e$} & \colhead{$l$} & \colhead{DM} &\colhead{EM} & \colhead{RM}\\
                         & ${\rm cm^{-3}}$ & pc & ${\rm cm^{-3}\,pc}$ & ${\rm cm^{-6}\,pc}$ & m$^{-2}$
}
\startdata
PDR & $10^5$ & ? & -- & -- & -- \\
IF & $>6000$ & $10^{-4}$ & $> 0.6$ & $> 3.6\times 10^3$ & $> 1.9\times 10^2$\\
Low Ion. & 7000 & $2\times 10^{-3}$ & 14 & $9.8\times 10^4$ &   $4.5\times 10 ^3$\\
Med Ion. & 4000 & 0.06 & 240 & $9.6\times 10^5$ & $7.7\times 10^4$
\enddata
\tablecomments{PDR: Photodissociation region. IF: Ionization Front. The temperature
and densities are from \citet{ODell01}.}
\label{tab:Orion}
\end{deluxetable}

\subsection{A Nuclear Magnetar Model}

\citet{pc15}, motivated by the discovery of SGR\, J1745$-$2900,
have proposed that FRBs arise from magnetars located in the centers
of galaxies (hereafter, ``nuclear magnetars").  The DM and RM of
this object is, as expected, extra-ordinary: ${\rm 1650\,cm^{-3}\,pc}$
and ${\rm -67000\,m^{-2}}$, respectively \citep{sj13}.  If such a
large contribution is accepted to the measured DM then a 
consequence is a vast reduction\footnote{
In this model, there would almost be no relation
between the true distance, $d$ and the DM.}
in the inferred distances to FRBs.   

We are puzzled by the hypothesis of nuclear magnetars.
There is  nothing special about SGR\,J1745$-$2900 other than
its nuclear location. There are several
magnetars similar to SGR\,J1745$-$2900, a middle-aged and relatively
quiet object, in other locations of the Galaxy. Any nuclear magnetar model 
must state what makes nuclear magnetars special enough to become
the sole source  of FRBs. Next, the model  is silent on a key observational
demand -- is  the population of active nuclear magnetars able 
to account for the observed  sky rate of FRBs? 

 \subsection{Young Pulsars}
 
 \citet{csp15} have proposed a model in which young extra-galactic pulsars, still within their supernova remnants,
 produce bright radio pulses which we see as FRBs. The supernova remnants contribute significantly 
 to the DM. The typical distance to FRBs, in this framework, is 200\,Mpc.
 In the introduction to \S\ref{sec:LocaleOfFRBs} we  noted the physical limitation of this
 model: the giant pulses of pulsars are extremely short (nanosecond corresponding to length scales
 of a meter; \citealt{hkw+03}). 
In order to produce FRBs we need coherently  emitting regions which are a few light milliseconds
wide  ($10^{11}\,$cm) -- far in excess of the magnetosphere of young pulsars. Regardless of 
this theoretical issue the model is testable. The error circles of FRBs should contain nearby
($<200\,$Mpc) galaxies.

\subsection{Summing up}

In summary, the young magnetar model can  account for the impressive
volumetric rate of FRBs,  the energetics  and also neatly explain the circumstantial
clues: scattering tails  that are seen in most FRBs and 
the occasional FRB with  strong free-free absorption.

We make the following parenthetical remark.  We do not know what
specific attribute of magnetars leads to the phenomenon of giant flares. Above we have discussed
the only two known Galactic SGRs which have emitted hyper-giant flares. Both SGRs are associated with star
clusters. The hyper-giant flare of 5 March 1979 came from SGR\,0525$-$66. This source 
is associated with a rich star-forming region in the Large Magellanic Cloud \citep{khg+04}.
The nominal characteristic age of SGR\,0525$-$66 is 3.4\,kyr \citep{ok14}.
There are 14 magnetars (SGRs and AXPs) whose characteristic age is less than 10\,kyr
(see \citealt{ok14}).
Two magnetars, 1E1547.0$-$5048
CXOU\,J171405.7$-$38103, have characteristic ages less than 1\,kyr. Some magnetars
are found close to the center of their supernova remnant. So while we do not know what
makes only some magnetars emit giant flares we do know that almost all of them lie in ISM rich
regions.  
 
 Continuing this parenthetical remark we note that \citet{mls+15} claim that two
 FRBs but with different DMs came from the same region in the sky. We have no
 independent assessment of this claim but note that it is entirely possible that
 a single galaxy, particularly one which is undergoing vigorous star-formation, to be 
 capable of hosting two FRBs within a few years.
 The differing DMs could be due to differing contributions from the local star-forming
 regions in which the magnetars are located.
 After all, the Milky Way (including the Magellanic Clouds), a system with a modest star-formation rate of
 only $\approx
 1\,M_\odot\,{\rm yr^{-1}}$ \citep{rw10}, has hosted four  hyper-giant flares in less than four
 decades.
 So it is quite reasonable that a galaxy like M51 which has star-formation rate three times
 that of our Galaxy could be three times more productive in giant flares relative to our Galaxy.

\section{Ramifications of the young magnetar model}
\label{sec:Ramifications}

In this section we explore the ramifications of the young magnetar model.
\begin{enumerate}

\item
Young magnetars are found in star-forming regions of galaxies. Thus,
the simplest consequence is that FRBs are not denizens of intergalactic
space.  If so, models involving non-stellar origin  (e.g.\ exotic
models involving super-conducting strings such as proposed by
\citealt{Vachaspati08}) can be rejected.

\item  
Next, as noted in \S\ref{sec:VolumetricRate}, the translation of
$\mathcal{R}$ to $\Phi_{\rm FRB}$ is usually done assuming that the
distance to the FRB is given by $d_{\rm IGM}$.  However, if we
accept the central thesis of this paper -- that FRBs are related
to young SGRs, which in turn mark the richest star-forming regions
in our Galaxy, then we also have to accept that a substantial
contribution to the measured DM of an FRB could also arise in rich
star-forming regions in the host galaxy (see discuss
\S\ref{sec:YoungMagnetarModel}).  Let $f=d/d_{\rm IGM}$ where $d$
is the true distance.  It is likely that $f$ is significantly smaller
than unity.

The true FRB rate is 
	\begin{equation}
	\Phi_{\rm FRB}^\prime = \langle f^{-3}\rangle \Phi_{\rm FRB}
	\label{eq:fPhiFRB}
	\end{equation}
where $\langle ... \rangle$ stands for the population average. FRBs
which have small DM but exhibit scattering or absorption will
contribute disproportionately to the true rate.  Regardless, the
true volumetric rate of FRBs is larger than had been hitherto 
discussed.

Separately, the FRB all-sky rate (Equation~\ref{eq:FRBDailyRate})
is based on FRBs with $\mathcal{F}>2\,{\rm Jy\,ms}$.  Indeed, the
rate of {\it detected} FRBs is nominally twice this rate \citep{cpk+15}.
Thus, the volumetric rate of FRBs could be larger
than that estimated by Equation~\ref{eq:fPhiFRB}.  If these two
factors lead to large corrections then it may well be that we need
to invoke\footnote{The volumetric rate of Giant Flares (as opposed
to Hyper-Giant Flares (Equation~\ref{eq:GF}) is off by a factor of
67 relative to the standard FRB volumetric rate
(Equation~\ref{eq:PhiFRB}).} giant flares from SGRs to account for
the increased volumetric rate of FRBs. 

We cannot make $f$ arbitrarily small.   Let $l_*$ be the
``typical" distance to FRBs. The volumetric rate is the
$\propto \mathcal{R}/l_*^3$. We equate this to the volumetric
rate of Giant flares. With that we find  $l_*\approx 0.8\,$Gpc.
FRBs are still very impressive extra-galactic transients.

\item 
As can be gathered from the discussion leading to Equation~\ref{eq:taud},
scattering tails require a highly turbulent medium. From basic
considerations one expects significant EM to be associated with
these turbulent regions (see \S11 of K14). Clearly, the detected
sample of FRBs, with the exception of FRB\,121102, does not show
strong free-free absorption at 1.4\,GHz.  However, the free-free
optical depth scales as $\nu^{-2.1}$. Thus, a modest\footnote{Such
optical depths cannot be ruled out for the observed sample of FRBs,
given the SNR.} optical depth of $\tau_{\rm ff}\approx 0.5$ at
1.4\,GHz will result in significant suppression in the meter-wave
band.  This issue is worthy of further study by the developers of
FRB searches with CHIME (400--800\,MHz) and the refurbished Molonglo
Telescope (843\,MHz).

\item 
One event, FRB\,140514, shows detectable circular polarization
\citep{pbb+15}.  By analogy to the pulsar phenomena, we would expect
FRBs to also exhibit linear polarization. A magnetized ISM will rotate the
linear polarization vector by an angle
	\begin{equation}
		\theta(\lambda) =  {\rm RM}\lambda_m^2,
	\end{equation}
where $\lambda_m$ is the wavelength in meters, and RM is the rotation
measure, $0.81\int n_eBdl$ with $B$, the magnetic field (along the
line of sight)  in units of $\mu$G and $dl$ in units of pc; the
units of RM is ${\rm m}^{-2}$.

The young magnetar model predicts significant RM (hundreds m$^{-2}$)
because of the increased magnetic field strength in HII regions
(\S\ref{sec:Orion}). In the case of FRB\,121102, we expect a very
large RM ($\sim 10^3\,{\rm m^{-2}}$ or more) because the magnetic
field in dense structures is expected to be significantly above the
typical diffuse ISM value. Those FRBs which show scattering tails
are expected to exhibit RMs larger than those which show no scattering
tails.

Finally,  for particularly bright FRBs, it may be possible to see
the local H~I and even some molecular species (OH)  in absorption
in the spectrum of the FRB. It would be worthwhile to design FRB
hardware such that a high quality and reliably calibrated spectrum can
be recovered from the  raw data (post-discovery).

\item 
SGR flares are expected to be accompanied by highly
relativistic ejecta.  The isotropic $\gamma$-ray energy released
in a hyper-giant flare is $3\times 10^{46}\, {\rm erg}$. It is not
unreasonable to assume that the ejecta has a kinetic energy comparable
to the $\gamma$-ray energy release. In our model, in some cases,
the ISM is close to the magnetar and so we expect the ejecta to slam
into high density ISM.  The resulting shock(s) can be expected to
generate radio, X-ray and even very high energy photons that could
last days to weeks.  In contrast, the same explosion taking place
in the average diffuse ISM will undergo a shock but on a longer 
timescale (e.g. type Ia supernovae).

\item A consequence of having a significant contribution to the
observed DM from ISM structures in the vicinity of the FRB
means that FRBs are noisy probes of the electron density and magnetic
field in the IGM (cf.\   \citealt{mcquinn14}) and of cosmography
(cf.\ \citealt{zlw+14,glz14}).  Delicate tests of cosmography may
prove to be challenging and frustrating (see also \citet{pc15}).
A possible way to minimize the outliers is to obtain the broad-band
spectrum, from meter wave to centimeter wave, of FRBs.
The broad-band spectrum will
readily identify FRBs with large free-free absorption.
However, this would 
require that facilities spanning a decade of frequency either be
truly band-band or enjoy common view.

\end{enumerate}

But all of this is speculation.  A clear test (and vindication) of
the SGR model is the detection of an FRB contemporaneous with a
soft $\gamma$-ray burst detection.

\acknowledgements

We are grateful to Dr. Laura Spitler for providing us the beam
response files.  We thank the PTF collaboration and in particular M.\ Kasliwal
and D.\ Cook  for allowing us to
showcase the PTF H$\alpha$ data ahead of a formal data release.  We
would like to thank Glenn Jones for asking us to consider  Galactic
ISS as a possible explanation for the large spectral index.  SRK's
research, in part, is supported by a grant from the US National
Science Foundation.

\bibliographystyle{apj1b}
\bibliography{bibAFRB}

\appendix

\section{Signal-to-Noise Ratio of the Bolometric Fluence}
\label{sec:BolometricFluence}

Consider a time series of spectra sampled every $\delta t$. Let the
frequency range from $\nu_1$ to $\nu_2$. For instance, for the
Arecibo data stream, the frequency range is 1225--1525\,MHz (with
960 channels with equal width) and $\delta t=65.5\,\mu$s (S14).

Consider the simple case where all the frequency bins are added up
with the same weight.  For simplicity, assume that the FRB has a
width of exactly $\delta t$. The bolometric flux is then the sum
of the frequency channels.  We replace the sum by an integral and
find
	\begin{equation}
		\mathcal{F} =\delta t \int_{\nu_1}^{\nu_2} f(\nu)d\nu
	\end{equation}
where $f(\nu)$ is the flux density at frequency $\nu$. Assume that
$f(\nu)$ is described (in the mean; denoted below  by a power law
with spectral index $\alpha$. The mean value (denoted below by
$\langle ... \rangle$)  and the variance (denoted by $V$) of the
bolometric fluence is thus
	\begin{eqnarray}
	 \langle \mathcal{F}\rangle &=& a\delta t
	\int_{\nu_1}^{\nu_2}(\nu/\nu_2)^\alpha d\nu,\\ 
	V(\mathcal{F})
	&=& \sigma_1^2(\nu_2-\nu_1)\delta t^2 \end{eqnarray},
where $a$ is the flux density at $\nu_2$ and $\sigma_1^2$ is the
variance per unit frequency (and assumed to be independent of
frequency). The basic radiometric equation informs us that $\sigma_1^2
= (T_{\rm sys}/G)(2\delta t)^{-1}$. Thus, the signal-to-noise ratio
(SNR) of the bolometric fluence is
	\begin{eqnarray}
		\mathcal{S} &=&
		\frac{\int_{\nu_1}^{\nu_2}a(\nu/\nu_2)^\alpha
		d\nu}{\sigma_1\sqrt{\nu_2-\nu_1}}\cr 
		&=& \frac{a}{\sigma}
		\frac{\nu_2}{(\alpha+1)(\nu_2-\nu_1)}
		\Big[[1-(\nu_1/\nu_2)^{\alpha+1}\Big],
		\label{eq:SNR1}
	\end{eqnarray}
where $\sigma^2 = \sigma_1^2/(\nu_2-\nu_1)$ is the (traditional)
rms for the entire band.

\begin{figure}[htbp] 
   \centering
   \includegraphics[width=7in]{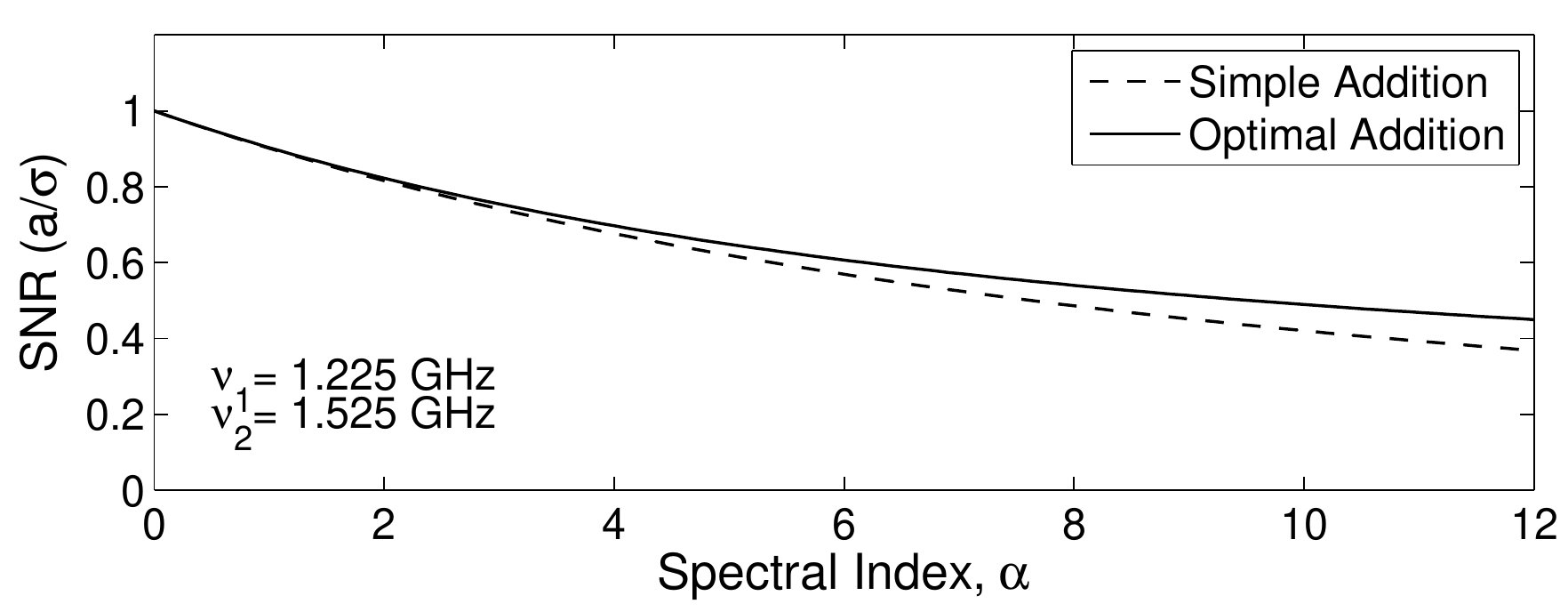}
   \caption{\small Signal-to-noise ratio of the bolometric fluence 
   of a pulse of radiation which has a power law spectrum (power
   law index, $\alpha$) over the frequency range $\nu_1=1225\,$MHz
   and $\nu_2=1525\,$MHz.    The dashed line is simple addition
   (Equation~\ref{eq:SNR1}) and the solid line is the performance of 
   optimal addition (Equation~\ref{eq:SNR2}).
	}
\label{fig:SNR12} 
\end{figure}

A flat spectrum source has a spectral index of 0 and thus
Equation~\ref{eq:SNR1} simplifies to $a/\sigma$, as expected. A
plot of the SNR as a function of $\alpha$ can be found in
Figure~\ref{fig:SNR12}. Notice the loss in SNR with increasing
$\alpha$.

\section{Optimal addition of channels}
\label{sec:OptimalAddition}

We wish to add the detected output from two frequency channels,
$x_1$ and $x_2$.  Let the statistics of $x_k$ ($k=1,2$) be each
described by independent Gaussian distribution with means of $\mu_k$
and variance, $\sigma_k^2$. Our goal is to find  a scheme that
maximizes the signal-to-noise ratio (SNR).

Let the optimal statistic be
	\begin{equation}
	X=\alpha x_1 + (1-\alpha)x_2.
	\end{equation}
Note the weights are already normalized, that is the sum of the two weights 
is unity.  The mean value, the variance, and the SNR of $X$ are given by
	\begin{eqnarray}
	\langle X\rangle &=& \alpha\mu_1 + (1-\alpha)\mu_2,\cr
	V(X) &=& \alpha^2\sigma_1^2 + (1-\alpha)^2\sigma_2^2,\cr   
	{\rm SNR} &=& \frac{\alpha\mu_1+(1-\alpha)\mu_2}{\sqrt{\alpha^2\sigma_1^2+(1-\alpha)^2\sigma_2^2}}.
	\label{eq:SNR}
	\end{eqnarray}

SNR is maximized by a judicious choice of $\alpha$. 
Differentiating Equation~\ref{eq:SNR} with respect
to $\alpha$ yields 
	\begin{equation}
		\alpha = \frac{\mu_1/\sigma_1^2}{\mu_2/\sigma_1^2+\mu_2/\sigma_2^2}.
	\end{equation}
The solution to this equation is  $\alpha=\mu_1/(\mu_1+\mu_2)$.
Thus, the weight is proportional to the expected signal strength
divided by the variance\footnote{This result is the basis of ``matched
filter''.}.  Substituting this weight into Equation~\ref{eq:SNR}
leads to the optimal SNR
	\begin{equation}
	{\rm SNR}=\sqrt{\frac{\mu_1^2}{\sigma_1^2}+\frac{\mu_2^2}{\sigma_2^2}}.
	\end{equation}
Thus, the optimal SNR is obtained when the SNRs of each measurement is added
in quadrature.

\section{Optimal Extraction of the Bolometric Fluence}

We now apply the results of \S\ref{sec:OptimalAddition} to optimally
extract the bolometric fluence. The symbols in this section have
the same meaning as those in \S\ref{sec:BolometricFluence}.

\begin{eqnarray}
	\mathcal{S}^2 &=&
		\Big(\frac{a}{\sigma_1}\Big)^2
		\Big[\int_{\nu_1}^{\nu_2}(\nu/\nu_2)^{2\alpha}d\nu\Big]\cr
		&=&
		\Big(\frac{a}{\sigma_1}\Big)^2
		\frac{\nu_2}{2\alpha+1}
		\Big[1-(\nu_1/\nu_2)^{2\alpha+1}\Big]\cr
		&=&
		\Big(\frac{a}{\sigma}\Big)^2
		\frac{\nu_2}{(2\alpha+1)(\nu_2-\nu_1)}
		\Big[1-(\nu_1/\nu_2)^{2\alpha+1}\Big].
		\label{eq:SNR2}
\end{eqnarray}
For the specific case of $\alpha=0$ we recover the result anticipated
on physical grounds, namely $\mathcal{S}=a/\sigma$. The performance
of the optimal addition as a function of $\alpha$ is shown in
Figure~\ref{fig:SNR12}. The optimal addition algorithm yields
moderately interesting gains only at large values of $\alpha$.

\section{Integrated H$\alpha$ Flux}
\label{sec:Halpha}

The IPHAS photometry, though employing Sloan bands and a custom
narrow band H$\alpha$ filter (hereafter, ``{\it ha}") is tied to
Vega magnitude.  We will adopt the following  zero points: $r^\prime$
(3012\,Jy), {\it ha} (3012\,Jy) and $i^{\prime}$ (2460\,Jy).  The
effective bandwidth of the H$\alpha$ filter is 95\,\AA\  and thus
the flux density integrated over the bandpass of the H$\alpha$ is
	\begin{equation}
	f({\rm H\alpha})= 2\times 10^{-15}10^{-0.4({\it ha}-20)}\,{\rm erg\,cm^{-2}\,s^{-1}}.
	\label{eq:HalphaFlux}
	\end{equation}
Alternatively, we can relate the bandpass integrated flux to the H$\alpha$ magnitude as 
follows:
	\begin{equation}
	{\it ha} = -2.5\log\Big[\frac{f({\rm H}\alpha)}{2\times 10^{-7}\,{\rm erg\,cm^{-2}\,s^{-1}}}\Big].
	\label{eq:HalphaMag}
	\end{equation}

\end{document}